\documentclass[nohyperref]{article}

\usepackage{hyperref}

\usepackage[accepted]{icml2022}

\usepackage{graphicx}
\usepackage{subcaption}
\usepackage{booktabs} 
\usepackage{cancel}
\usepackage{wrapfig}
\usepackage{placeins}
\usepackage{lipsum}
\usepackage{yfonts}
\usepackage{bm}
    
\usepackage{pifont} 

\usepackage{hyperref} 
\hypersetup{colorlinks,linkcolor={blue},citecolor={blue},urlcolor={blue}}

\usepackage{nicefrac}       
\usepackage{amssymb,amsfonts}       
\usepackage{amsmath,stackengine,mathtools}
\usepackage{amsthm}
\usepackage{tcolorbox}

\makeatletter
\def\moverlay{\mathpalette\mov@rlay}
\def\mov@rlay#1#2{\leavevmode\vtop{%
   \baselineskip\z@skip \lineskiplimit-\maxdimen
   \ialign{\hfil$\m@th#1##$\hfil\cr#2\crcr}}}
\newcommand{\charfusion}[3][\mathord]{
    #1{\ifx#1\mathop\vphantom{#2}\fi
        \mathpalette\mov@rlay{#2\cr#3}
      }
    \ifx#1\mathop\expandafter\displaylimits\fi}
\makeatother

\usepackage[utf8]{inputenc}

\usepackage{xspace}
\usepackage[T1]{fontenc}    
\usepackage{cleveref}
\usepackage{diagbox}
\crefformat{section}{\S#2#1#3}
\crefformat{subsection}{\S#2#1#3}
\crefformat{subsubsection}{\S#2#1#3}

\theoremstyle{definition}
\newtheorem{definition}{Definition}

\theoremstyle{definition}

\theoremstyle{definition}

\theoremstyle{definition}

\theoremstyle{definition}

\crefname{axiom}{ax.}{axs.}
\Crefname{axiom}{Axiom}{Axioms}
\creflabelformat{axiom}{#2#1#3}
\crefname{definition}{def.}{defs.}
\Crefname{definition}{Definition}{Definitions}
\creflabelformat{definition}{#2#1#3}
\crefname{proposition}{prop.}{props.}
\Crefname{proposition}{Proposition}{Propositions}
\creflabelformat{proposition}{#2#1#3}
\crefname{remark}{remark}{remarks}
\Crefname{remark}{Remark}{Remarks}
\creflabelformat{remark}{#2#1#3}

\usepackage{todonotes}

\usepackage{enumitem}
\setlist[itemize]{leftmargin=*}
\usepackage{multirow}
\usepackage{dsfont}

\usepackage{tikz}
\usetikzlibrary{shapes,decorations,arrows,calc,arrows.meta,fit,positioning}
\tikzset{
    -Latex,auto,node distance =1 cm and 1 cm,semithick,
    state/.style ={ellipse, draw, minimum width = 0.7 cm},
    point/.style = {circle, draw, inner sep=0.04cm,fill,node contents={}},
    bidirected/.style={Latex-Latex,dashed},
    el/.style = {inner sep=2pt, align=left, sloped}
}
\DeclareMathOperator*{\argmin}{arg\,min}

\newcommand{\mat}[1]{\mathbf{#1}}
\renewcommand{\vec}[1]{\mathbf{#1}} 

\newcommand{\expectation}[2]{ \mathbb{E}_{#1}{\left[#2\right]} }
\newcommand{\x}{\vec{x}}
\newcommand{\X}{\vec{X}}

\newcommand{\Y}{\vec{Y}}

\newcommand{\V}{\vec{V}}

\newcommand{\graph}{\mathcal{G}}

\newcommand{\Vnet}{\V_{\textrm{net}}}

\newcommand{\dataset}{\mathcal{D}}

\newcommand{\DObs}{\dataset^\text{O}}

\newcommand{\DO}[2]{\operatorname{do} \!  \left(#1 = #2\right)}

\newcommand{\scmdef}{\left\langle \mat{U},\mat{V},\mat{F}, p(\mat{U}) \right\rangle}

\newcommand{\dom}[1]{\text{dom}(#1)}

\newcommand{\allX}{\bm{\mathcal{X}}_{t-1}}
\newcommand{\patch}{\textrm{RES}}
\newcommand{\isolate}{\textrm{ISO}}
\newcommand{\vuln}[1]{\textrm{vuln}(#1)}

\newcommand{\nd}[1]{{\color{red}[Neil: #1]}}

\newcommand{\al}[1]{{\color{orange}[Alex: #1]}}

\newcommand{\acro}[1]{\textsc{#1}\xspace}

\newcommand{\DAG}{\acro{dag}}

\newcommand{\dcbo}{\acro{dcbo}}
\newcommand{\cbo}{\acro{cbo}}
\newcommand{\gp}{\acro{gp}}
\newcommand{\bo}{\acro{bo}}

\newcommand{\dcgo}{\acro{dcgo}}

\newcommand{\hvt}{\acro{hvt}}
\newcommand{\rl}{\acro{rl}}
\newcommand{\ppo}{\acro{ppo}}

\newcommand{\cbs}{\acro{cbs}}

\newcommand{\scm}{\acro{scm}}
\newcommand{\sem}{\acro{sem}}
\newcommand{\yt}{\acro{yt}}
\newcommand{\ba}{\acro{ba}}
\newcommand{\ra}{\acro{ra}}
\newcommand{\acd}{\acro{acd}}

\icmltitlerunning{Optimal Causal Cyber-Defence Agents}

\begin{document}

\twocolumn[
\icmltitle{Developing Optimal Causal Cyber-Defence Agents \\ via Cyber Security Simulation}



\icmlsetsymbol{equal}{*}

\begin{icmlauthorlist}
\icmlauthor{Alex Andrew}{equal,comp}
\icmlauthor{Sam Spillard}{equal,yyy}
\icmlauthor{Joshua Collyer}{comp}
\icmlauthor{Neil Dhir}{yyy}
\end{icmlauthorlist}

\icmlaffiliation{yyy}{Defence \& Security Group, The Alan Turing Institute, London, UK}
\icmlaffiliation{comp}{Defence Science and Technology Laboratory, Salisbury, UK}

\icmlcorrespondingauthor{Sam Spillard}{sam.s@turing.ac.uk}
\icmlcorrespondingauthor{Neil Dhir}{ndhir@turing.ac.uk}

\icmlkeywords{Machine Learning, ICML}

\vskip 0.3in
]



\printAffiliationsAndNotice{\icmlEqualContribution} 

\begin{abstract}
    In this paper we explore cyber security defence, through the unification of a novel cyber security simulator with models for (causal) decision-making through optimisation. Particular attention is paid to a recently published approach: dynamic causal Bayesian optimisation \citep[\dcbo]{dcbo}. We propose that \dcbo can act as a blue agent when provided with a view of a simulated network and a causal model of how a red agent spreads within that network. To investigate how \dcbo can perform optimal interventions on host nodes, in order to reduce the cost of intrusions caused by the red agent. Through this we demonstrate a complete cyber-simulation system, which we use to generate observational data for \dcbo and provide numerical quantitative results which lay the foundations for future work in this space.
\end{abstract}

\section{Introduction}
\label{sec:intro}

In a recent paper \citet{dhir2021prospective} voice concern regarding cyber criminals developing new and complex malicious tools which defenders cannot rapidly adapt to, or rapidly counter with existing methods. They argue that the sophistication of these new tools warrants the entry of new advanced cyber-defence techniques to counter these threats. For the purposes of active cyber defence (\acd), they propose the rapid introduction of reinforcement learning (\rl) and causal inference. We focus on the latter in this paper due to the much larger, existing, body of work that has hitherto focused on \rl for cyber security applications see e.g. \citep{ridley2018machine} and \citep{sewak2021deep} for a review. 

The actions that a blue agent (\ba) takes when acting as a defender of a network node can be viewed as a solution to an optimal decision making problem. To have the best chance of survival at each time step, the \ba acting as a defender must make the optimal intervention to protect itself. The \ba is effectively trying to minimise the probability that it will be compromised, given the current state of the network. This becomes particularly complex in the computer network setting, as the causal relationships between network properties and the chance that a node will be compromised vary over time, as a red agent (\ra) takes sequential malicious actions on the network.

To find the optimal interventions which the \ba should take to counter the actions of the \ra, we propose the use of \dcbo which was introduced to identify optimal interventions in precisely this kind of setting. Furthermore, it was shown to converge faster than competing methods in a variety of similar dynamic scenarios \citep{dcbo}. \dcbo is able to provide the optimal sequence of interventions, accounting for causal temporal dynamics, in a dynamical system such as that of a network under attack. This optimisation is performed in relation to a specific cost function, such as minimising network down-time or a per-node appraisal. In order to investigate interventional dynamics, we apply \dcbo to data simulated by Yawning Titan (\yt) -- a novel cyber security simulator developed by the UK Government to test new solutions to existing problems within the cyber security domain.



For our simulations, the \ba has two available actions. They can either (1) restore a compromised node (thereby removing \ra from that node) or (2) isolate the node (removing the connections from that node to other nodes, thereby making it impossible for the \ra to reach it) for a number of time steps. In order to avoid trivial solutions in this setting, we impose a cost to both actions, as well as a cost to nodes in the network being compromised. The optimal set of interventions will then be those that minimise the total cost to the agent -- keeping the network free from \ra infection while minimising the use of expensive actions.

The paper is organised as follows; in \cref{sec:prelims_problem_setup} we review some necessary background material for the optimisation problem. This leads into related work on cyber simulators in \cref{sec:related_work}, allowing us to introduce \yt in \cref{sec:yawning_titan}. Having gone through \yt, we demonstrate how we use it with models for (causal) decision-making through optimisation, with our methodology in \cref{sec:methodology}. We present experiments, results and conclusion in \cref{sec:experiments}, \cref{sec:results_n_discussion} and \cref{sec:conclusion} respectively.
\section{Preliminaries and problem setup}
\label{sec:prelims_problem_setup}

\paragraph{Notation.} Random variables are denoted by upper-case letters and their values by lower-case letters. Sets of variables and their values are noted by bold upper-case and lower-case letters respectively. We make extensive use of the do-calculus, for details see \citep[\S 3.4]{pearl2000causality}. Samples (observational) drawn from a system or process unperturbed over time are contained in $\mathcal{D}^O$. 
\paragraph{Structural causal model.} Structural causal models (\scm{}s) \citep[ch. 7]{pearl2000causality} are used as the semantics framework to represent an underlying environment. For the exact definition as used by Pearl see \citep[def. 7.1.1]{pearl2000causality}. An \scm is parametrised by the quadruple $\scmdef$. Here, $\mat{U}$ is a set of exogenous variables which follow a joint distribution $p(\mat{U})$ and $\mat{V}$ is a set of endogenous (observed) variables. Within $\mat{V}$ we distinguish between three types of variables: $\X$, manipulative (treatment); $\mathbf{Z}$, non-manipulative covariates and $\Y$ the target (output) variables. The sets of parents, children, ancestors, and descendants in a graph $\graph$ will be denoted by PA, CH, AN and DE respectively.

Further, endogenous variables are determined by a set of functions $\mat{F} \subset \mathcal{F}$. Let $\mat{F} \triangleq \{f_i\}_{V_i \in \mat{V}}$ \citep[\S 1]{lee2020characterizing} s.t. each $f_i$ is a mapping from (the respective domains of) $U_i \cup \textrm{PA}_i$ to $V_i$ -- where $U_i \subseteq \mat{U}$ and $\textrm{PA}_i \subseteq \mat{V} \setminus V_i$. Graphically, each \scm{} is associated with a causal diagram (a directed acyclical graph, \DAG for short) $\graph  = \left\langle \mat{V}, \mat{E} \right\rangle$ where the edges are given by $\mat{E}$. Each vertex in the graph corresponds to a variable and the directed edges point from members of $\textrm{PA}_i$ and $U_i$ toward $V_i$ \citep[ch. 7]{pearl2000causality}. A directed edge is s.t. $V_i \leftarrow V_j \in \mat{E}$ if $V_i \in \textrm{PA}_j$ (i.e. $V_j$ is a child of $V_i$). 

\subsection{Problem statement}

Formally we can pose our problem setting as version of the \emph{control problem} as defined by \citet[\S 2]{pearl1995probabilistic}. The setting consists of a \DAG, associated vertex set $\V$ partitioned into four disjoint sets $\V = \{\X, \mathbf{Z}, \mathbf{U}, Y\}$ \emph{for each} time-slice. There exists a temporal order on the disjoint sets so that $\V = \bigcup_{t=0}^T \V_t$ so that each slice is an ancestor of $\mathds{1}_{t>0} \cdot \V_{t}$ in $\graph$. Each slice contains \emph{one} outcome variable with index $t$. Each slice is a sub-graph of $\graph$ and each sub-graph is a directed rooted tree where the root node is the response variable $Y_t$. In the offline setting we are privy to observational samples contained in $\DObs$ and from that information we seek a \emph{plan} which is an ordered sequence of interventions which when implemented, jointly maximise (minimise) the associated sequence of response variables $\{Y_0,Y_1,\ldots, Y_T\}$. 

Using causal inference notation we seek the interventional plan: $ \{\DO{\X^*_0}{\x^*_0},\DO{\X^*_1}{\x^*_1},\ldots,\DO{\X^*_T}{\x^*_T}\}$ which is found by solving a  \emph{dynamic causal global optimisation} \citep{dcbo} problem of the form:
\begin{align}
    \label{eq:dcgo}
    \X_t^*,\x_t^* 
    &= \argmin_{\X_t \in \mathcal{P}(\X_t), \x_t \in \dom{\X_t}}  \nonumber \\
    &\expectation{}{\Y_t \mid \DO{\X_t}{\x_t}, \mathds{1}_{t>0} \cdot \allX}
\end{align}
where $\allX \triangleq \bigcup_{i=0}^{t-1} \DO{\X_{i}}{\x_{i}}$ and $\mathcal{P}(\cdot)$ is the powerset over the interventional variables at time $t$.

This setting makes a number of assumptions which are required to a sequence of optimal actions within a causal framework (the details of which can be found in \citep[Assumptions 1]{dcbo}). Notwithstanding, we seek, at every time step, to construct models for different intervention sets (the expectation in \cref{eq:dcgo}) by integrating various sources of data while accounting for past interventions. With this construction we are able to query them to find an optimal intervention at each time-index given the history.



\section{Related work}
\label{sec:related_work}
As noted in \cref{sec:intro} much of the literature has focused on the use of \rl in cyber security. We will instead consider relevant literature on cyber simulators (\cref{sec:cyber_simulators}) and the few studies which have used concepts from causal inference for cyber security research (\cref{sec:causal_cyber}).
\subsection{Cyber simulators}
\label{sec:cyber_simulators}
There are many cyber simulators that exist or are currently in development, so why the need for \yt in this work? We consider our contemporaries to motivate the use of \yt.

FARLAND \citep{molina2021network} aims to support the development of \rl agents within a hybrid simulation and emulation system. This increased fidelity provides a means of developing \rl cyber defence agents capable of defending real systems but hinders the research of fundamental aspects such as algorithm development. \yt is more abstract and flexible which enables us to easily test different and novel approaches without having to significantly re-engineer the underlying code base. CybORG \citep{standen2021cyborg}, comparatively, shares many features with \yt. Both are simulators and are controlled by \texttt{YAML} configuration files that define the scenario for a given environment. The fundamental difference however is the fidelity of node representation. As CybORG contains both a simulation and emulation component, the simulation models real features of hosts, such as the operating systems, hardware architectures, users and passwords. All of this information is intentionally abstracted away within \yt. This level of abstraction could be viewed negatively but is the very reason new and novel decision making approaches can be rapidly integrated and experimented with whilst also providing a cyber relevant environment. Another recent development is Microsoft's Cyber Battle Simulator \citep[\cbs]{msft:cyberbattlesim}.  Similar to \yt, \cbs abstracts away from many of the details of real life however still incorporates host information such a specific host vulnerabilities and in-depth environment setup. When designing \yt we wanted to focus on a very simple setup where the hosts only have a few shared attributes and it is easy to spin up new network models. This would allow someone to quickly set up an abstraction of a network topology, without having to worry about host-specific details, and easily integrate it into their own research.

\subsection{Causality for cyber security}
\label{sec:causal_cyber}

Causal inference methods have yet to find their place within the cyber security domain. There have been some previous attempts to apply causal decision making to historic cyber data by \citet{ApplicationsCausalCyber} and \citet{8999155}. However, these attempts use causal models without a temporal dimension (the main feature of \dcbo) and only focus on predicting and alerting rather than selecting optimal defensive actions. Since we are using a simulator to collect our data, we can easily test new scenarios and model performance differences that one would not be able to with a single historic data set.

In another strand of work, \citet{shi2017causality} use transfer entropy to quantify the causal relationship between any two variables in their cyber-physical system. In other words they undertake causal-discovery of their complex network in order to not a priori have to rely on a model of the underlying dynamical system.

\section{Yawning titan}
\label{sec:yawning_titan}

We introduce Yawning Titan (\yt). \yt is an abstract, highly flexible, cyber security simulator that is capable of simulating a range of cyber security scenarios\footnote{A Python implementation is available: \url{https://github.com/dstl/YAWNING-TITAN}.}. We provide a formal definition for completeness.
\begin{definition}{Cyber security simulator.}
Simulation in the cyber context is the process of modelling a real-world computer-system environment to predict outcomes of actions from a number of agents where the goals of said agents are to control either system or data from the system.
\end{definition}
There are a number of these simulators (see \cref{sec:related_work} for related work), who broadly work on the same principle as described in the above definition. \yt is built to train cyber defence agents to defend arbitrary network topologies in highly customisable configurations. It was developed with the following design principles in mind: 
\begin{enumerate}[noitemsep,topsep=0pt]
    \item Simplicity over complexity
    \item Minimal hardware requirements
    \item Support for a wide range of algorithms
    \item Enhanced agent/policy evaluation support
    \item Flexible environment and game-rule setup
\end{enumerate}
Design principles 1, 2 and 5 make \yt an ideal environment to test and evaluate new and novel approaches to cyber security decision making and provides a means of transitioning approaches such as \dcbo into a cyber defence context. Principally we will be modelling the actions of one blue agent (\ba) and one red agent (\ra).

\begin{figure*}[ht!]
    \centerline{\includegraphics[width=\textwidth]{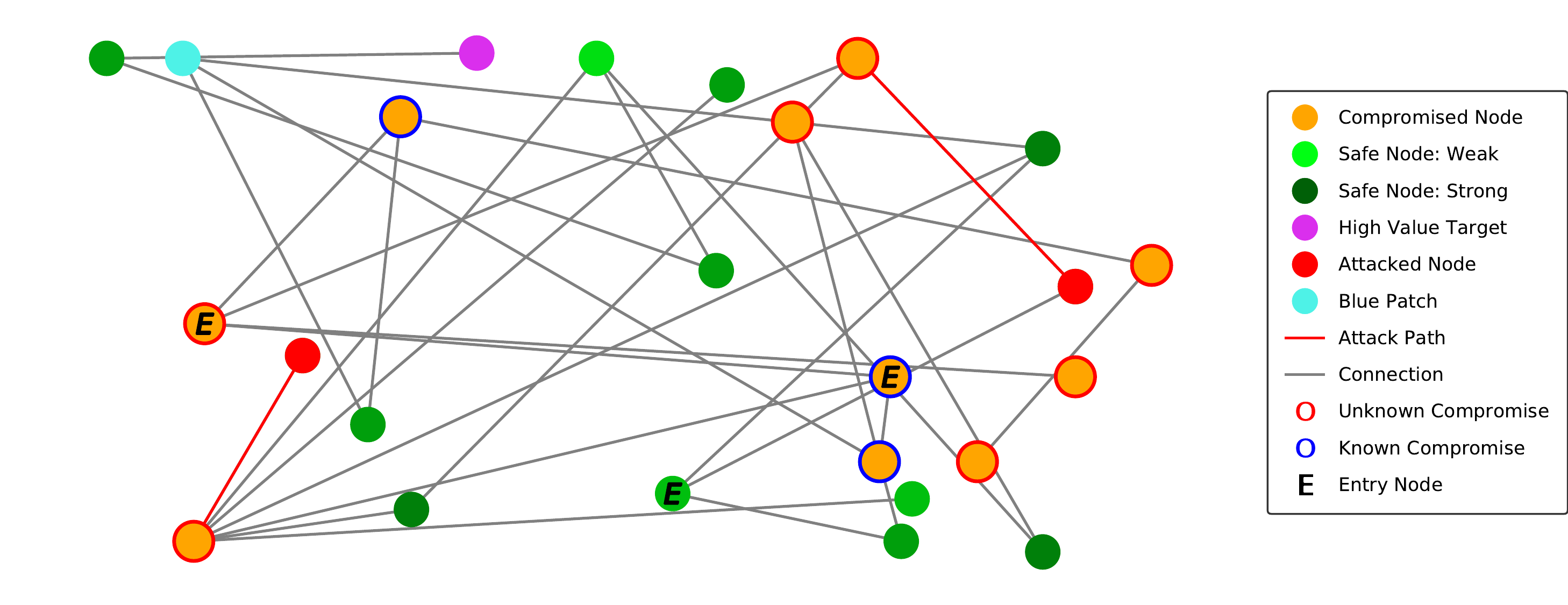}}
    \caption{Example output from \yt on a 25 node network $\graph_{\textrm{net}}$. The \ba is defending a high value target node (purple node). The \ra controls a large portion of the network, indicated by yellow nodes with known and unknown compromise (indicated by the edge colour of the yellow nodes), unbeknownst to the \ba. There are three entry nodes into the system, two of which the \ra has made use of (marked by an `E'). The \ba is in the process of removing the \ra from the node adjacent to the high value target node. The \ra is attacking a node in the bottom left hand corner of the network. Best viewed in colour.}
    \label{fig:YT_simulation}
\end{figure*}

Let $\graph_{\textrm{net}}  = \left\langle \Vnet, \mat{E}_{\textrm{net}} \right\rangle$, with vertex set $\Vnet$ and edge set $\mat{E}_{\textrm{net}}$, be the undirected graph which simulates a computer network in \yt. Before going further, take care not to confuse the two types of graphs we employ in this work; $\graph$ represents the directed causal diagram and $\graph_{\textrm{net}}$ the undirected associated computer network -- an example instance is shown in \cref{fig:YT_simulation}. Each node $V_i \in \Vnet$ represent the hosts on the network and arc $E_j \in \mat{E}_{\textrm{net}}$ represents possible connectivity between the two connected hosts (grey edges in \cref{fig:YT_simulation}). This approach allows us to easily model a large variety of different types of network topologies in their simplest form without worrying about protocols or different types of hosts. This basic approach to the network allows us to abide by our first design principle.

\yt allows for considerable customisation through the use of customised configurations. The configuration summary contains a number of different settings that the user can control in order to change how the game is played. From these settings you can change success conditions for the agents and how each of the different agents can act by toggling actions and action probabilities.

The machines in the network each have their own attributes that affect how they behave and how they are affected by the \ra and the \ba.
\begin{description}
   \item[Vulnerability score] Affects how easy it is for the \ra to compromise the node. It is automatically generated at the start of each scenario based on settings in the configuration file and there are blue actions that can modify the vulnerability scores of nodes.
   \item[Isolation Status] Indicates whether a node has been isolated. This means that all of the incoming and outgoing connections are disabled. This can be modified by blue actions.
   \item[True Compromised Status] Indicates whether a node has been infected by the \ra. If the \ra has control of the node then it can use the node as a foothold to spread to other connected nodes. 
   \item[Blue seen Compromised Status] Represents if the node is compromised and the \ba is aware of the intrusion. Depending on the scenarios configuration, the \ba may only see an obscured view of the network and see this instead of the true value, effectively simulating the differences between perfect and imperfect detection.
\end{description}
Using the aforementioned configuration settings, the user is able to modify the subset of actions that are available to \ra and \ba. The \ba has actions that allow it to: 
\begin{itemize}[noitemsep,topsep=0pt]
    \item reduce the vulnerability of nodes, 
    \item scan the network for red intrusions, 
    \item remove \ra from a node, 
    \item reset a node back to its initial state, 
    \item deploy deceptive nodes, 
    \item isolate a node and 
    \item reconnect a previously isolated node.
\end{itemize}
The \ra has a variety of different attacks that it can use based on the settings in the configuration file. Unless using a guaranteed attack, \cref{eq:basic_attack_1} will be used to determine if the \ra (with a skill level $0 < RS \leq 1$) will compromise node $V_i \in \V_{\textrm{net}}$:
\begin{equation}
    \label{eq:basic_attack_1}
    AS  = \frac{100 \times RS^2}{RS + (1 - \vuln{V_i})}.
\end{equation}
The attack will succeed if:
\begin{equation}
    \label{eq:basic_attack_2}
   AS \geq u 
\end{equation}
where $u \sim \mathcal{U}(0,100)$ is sampled from a uniform distribution, $AS$ is the attack score (a measure of how powerful an attack is) and $\vuln{V_i}$ is the vulnerability of node $V_i$ (a measure of how exposed a host is to being compromised). For example, a computer without a firewall or a user that has no security training, could be modelled as having high vulnerability. Since $0 \leq \vuln{V_i} \leq 1$ and $0 \leq AS \leq 100$, the use of this formula ensures that the likelihood for the \ra to compromise a node increases proportionately with the skill of the agent and decreases proportionately with the defence of the node $(1 - \vuln{V_i})$. 

At each time step, each agent is allowed one action, chosen from the subset of activated actions, to affect the environment. The order of \yt execution is as follows:
\begin{enumerate}[noitemsep,topsep=0pt]
    \item The \ra performs their action
    \item The Environment checks if the \ra has won
    \item The \ba performs their action
    \item The Environment returns the reward of the \ba's action
    \item The Environment checks if the \ba has won.
\end{enumerate}
As \yt is built upon the \texttt{OpenAI Gym} framework \citep{1606.01540}, the code is simple to understand and highly decoupled. The agents can be any function that picks an action based on an observation matrix. Consequently, although \yt was built to run with a \texttt{Stable-Baselines3} \citep{stable-baselines3} it is simple to model the actions of the \ba using a causal sequential decision-making agent.

\section{Methodology}
\label{sec:methodology}

In this section we characterise the various building-blocks required for the optimal intervention-recommendation part of this study. First and foremost, we consider the different decision-making methods under investigation -- for continuous action spaces. All are based on Bayesian optimisation \citep[\bo]{movckus1975bayesian,garnett_bayesoptbook_2022} which is our weapon of choice to solve \cref{eq:dcgo}.

\bo is an optimisation method used for global optimisation of black-box functions \citep[\S 1]{garnett_bayesoptbook_2022} -- i.e. those which do not assume a specific functional form \citep{mockus2012bayesian} and are only distinguished by their inputs and outputs. It finds use in numerous domains \citep{garnett_bayesoptbook_2022}, but is principally used for optimising expensive black-box functions (where the cost of an evaluation can be e.g. monetary, time-dependent or societal). \bo has key components:
\begin{enumerate}
    \item the objective function is modelled with a Gaussian process (\gp) \citep{williams2006gaussian} -- the expectation in \cref{eq:dcgo}; 
    \item each new evaluation of the objective function is incorporated via a Bayesian update procedure and 
    \item an acquisition function is used to determine the next, high utility, point of evaluation of the objective. 
\end{enumerate}
Numerous advancements and improvements to \bo have been made over the years, but here we focus on those that extend the original framework to the explicitly causal setting. The first causal extension to \bo is given by the causal Bayesian optimisation (\cbo) model, introduced by \citet{cbo}. \cbo is used in settings where the response variable $Y$, is part of a \scm in which a sequence of interventions can be performed. \cbo is designed for \emph{static} environments. It does not account for the temporal evolution of the system, consequently breaking the temporal dependency structure which exists among variables -- see \cref{fig:DAG} for the causal diagram used in this study. 

In order to handle problem spaces with explicit temporal dynamics, \citet{dcbo} introduced dynamic causal Bayesian optimisation (\dcbo). \dcbo is useful in the kind of scenarios considered in this paper; ones where all causal effects in the causal diagram are changing over time. At every time step \dcbo identifies a local optimal intervention by integrating both observational and past interventional data collected from the system. Precisely, a \dcbo agent seeks to minimise the objective function while accounting for the cost of intervention. It is important to note that interventions typically have a large financial, societal, ethical or other cost associated with them. This is true in the cyber setting as well where a potentially optimal intervention could be e.g. shutting down a node but which has huge financial cost due to the services running on that node. Hence an agent needs to optimise the objective function whilst taking \emph{that} interventional cost into account.



Using observational data from \yt and the three methods described above, we investigate optimal action policies which reduce and prevent intrusions into a network from a \ra, by solving the \dcgo problem in \cref{eq:dcgo}. We now outline relevant causal components.





\subsection{Causal diagram} 
\begin{figure}[ht!]
    \centering
        \begin{tikzpicture}[node distance =0.75cm and 2cm]
            
            \node (P0) [label=right:{$P_0$},point];
            \node (I0) [label=above:{$I_0$},below of = P0,point];
            \node (S0) [label=above right:{$S_0$},below of = I0,point];
            \node (C0) [label=above:{$C_0$},below of = S0,point];
            \node (H0) [label=below:{$H_0$},below of = C0,point];
            \node (A0) [label=right:{$A_0$},below of = H0,point];
            \node (T0) [label=right:{$T_0$},below of = A0,point];
       
            \path (I0) edge (S0);
            \path (C0) edge (H0);
            \path (A0) edge (T0);
            \path (P0) edge[bend right=50] (H0);
            \path (P0) edge[bend right=50] (A0);
            \path (I0) edge[bend right=50] (A0);
            \path (C0) edge[bend right=50] (T0);
            
            \node (P1) [label=right:{$P_1$}, right = of P0, point];
            \node (I1) [label=above:{$I_1$},below of = P1,point];
            \node (S1) [label=above right:{$S_1$},below of = I1,point];
            \node (C1) [label=above:{$C_1$},below of = S1,point];
            \node (H1) [label=below:{$H_1$},below of = C1,point];
            \node (A1) [label=right:{$A_1$},below of = H1,point];
            \node (T1) [label=right:{$T_1$},below of = A1,point];
            
            \path (H0) edge (C1);
            \path (C0) edge (S1);
        
            \path (I1) edge (S1);
            \path (C1) edge (H1);
            \path (A1) edge (T1);
            \path (P1) edge[bend right=50] (H1);
            \path (P1) edge[bend right=50] (A1);
            \path (I1) edge[bend right=50] (A1);
            \path (C1) edge[bend right=50] (T1);
            
            \node (P2) [label=left:{$P_2$},right = of P1,point];
            \node (I2) [label=above:{$I_2$},below of = P2,point];
            \node (S2) [label=above left:{$S_2$},below of = I2,point];
            \node (C2) [label=above:{$C_2$},below of = S2,point];
            \node (H2) [label=below:{$H_2$},below of = C2,point];
            \node (A2) [label=left:{$A_2$},below of = H2,point];
            \node (T2) [label=left:{$T_2$},below of = A2,point];
            
            \path (C1) edge (S2);
            \path (H1) edge (C2);
            
            \path (I2) edge (S2);
            \path (C2) edge (H2);
            \path (A2) edge (T2);
            \path (P2) edge[bend left=50] (H2);
            \path (P2) edge[bend left=50] (A2);
            \path (I2) edge[bend left=50] (A2);
            \path (C2) edge[bend left=50] (T2);
            \end{tikzpicture}
    \caption{Causal diagram $\graph$ shown for the first three time-steps. The target variable is $T$ with manipulative variables given by $\X_t = \{P_t,I_t\}$. The rest are non-manipulative variables. The within slice topology repeats for 25 time-steps in our experimental setup and the variable connectivity repeats as shown for the same length of time. For variable descriptions see \cref{table:variable_descriptions}.} 
    \label{fig:DAG}
\end{figure}
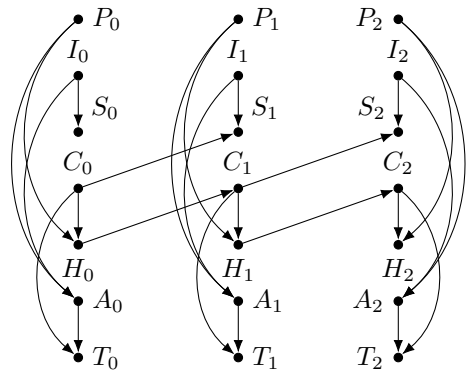

The causal diagram induced by the \scm{} is shown in \cref{fig:DAG} and is in accordance with causal sufficiency (no unobserved confounders) as per the assumptions made in \citep[assumptions 1]{dcbo}. Further, the graph topology is homogeneous within a time-slice and remains so across time. Simply, the \DAG prescribes the causal relationships between endogenous variables within and across time. Variable descriptions are given in \cref{table:variable_descriptions}.

\begin{table*}[ht!]
    \centering
    \caption{\scm variable descriptions. For their causal relationship see \cref{fig:DAG}.}
    \label{table:variable_descriptions}
    \vspace{-1em}
    \begin{tabular}{p{0.075\linewidth}p{0.75\linewidth}}
    \toprule
    Variable & Description  \\
    \midrule
    $P$      & Probability that a \ba will restore (\patch) a node, thereby removing the \ra from that node.\\
    $I$      & Probability that a \ba will isolate (\isolate) a node, thereby making both inbound and outbound attacks from a \ra impossible.\\
    $S$      & Available attack surface for the \ra. Number of nodes that are connected to the network and not yet compromised.\\
    $C$      & Operating cost of a node being compromised. We assume that all nodes have equal operating cost of compromise.\\
    $H$      & Likelihood of further compromise. This is equal to the total vulnerability of all nodes that are directly connected to any compromised node.\\
    $A$      & Operating cost of taking a given action. We define different costs for both \patch{} and \isolate.\\
    $T$      & Total operating cost. Sum of $C$ and $A$.\\
    \bottomrule
    \end{tabular}
\end{table*}


This \DAG is chosen as we believe it represents the simplest causal structure of how compromise spreads through a network, whilst accounting for the specific operating costs of taking actions to prevent further intrusion. It is important to factor in these costs to avoid trivial solutions, such as disconnecting every node from the network (despite that sometimes being the best option in a hostile cyber environment)! Additionally, in order to ensure the temporal functional form of the \ra is captured in our model, we choose to explicitly dynamically model $S$ and $H$, the attack surface and likelihood of additional compromise. This allows the results of actions that the \ra takes in between time steps, as well as the causal relationships, to be explicitly modelled. Due to our total control of the \yt environment, we are able to make certain assumptions that would not hold in an equivalent real-world setting, such as; the only \ba actions being taken are \patch\ or \isolate, and only one at a time; the \ra can only spread through the network from nodes it already controls, and cannot create additional entry points; and the topology of the network remains consistent across time, apart from those cases where a node is disconnected by the \ba. These are strong assumptions but critical to ensure our \DAG accurately represents the data generating process.

\subsection{Structural equation model} 

The \DAG is chosen to represent the most important properties of the environment that could also be used in a real world scenario. The nodes represent the chance that a system can become compromised, how that compromise spreads, and costs associated with that compromise.

We setup \yt as a data-generator in order to collect observational data, used to fit the causal relationships prescribed by the directed edges in $\graph$ in \cref{fig:DAG}. For this, we initialise a \ba with a two-dimensional action space of $\{ \text{restore } (\patch), \text{isolate } (\isolate)\}$ and random action probabilities -- full description in \cref{table:variable_descriptions}. 

The functional relationship between variables, the true \sem, are provided in \cref{eq:sem} to \cref{eq:sem_finish}. Here $K_t$ is the subset of all nodes that are compromised at time $t$ and $\phi_t$ the subset of all nodes that are isolated at time $t$. Where $\phi^{c}$ denotes the complement of subset $\phi$, $N^{+}(V_i)$ denotes all nodes that can be reached via a single edge from node $V_i$ (see \cref{fig:YT_simulation} for an example), $\Gamma_{\{c,\ \patch,\ \isolate \}}$ represents the cost of compromise, restore and isolate respectively, $\mathcal{A}_t$ represents the action the \ba took at time $t$, $\vuln{V_i}$ represents the vulnerability of node $V_i$ and $[\cdot]$ represents the Iverson bracket. The exponent in \cref{eq:C} is implemented in order to improve convergence of an \rl agent trained on a similar environment, so we include it here for consistency.
\begin{align}
    P_t &= p_{t}(\mathcal{\patch}) \label{eq:sem} \\
    I_t &= p_{t}(\mathcal{\isolate}) \\
    S_t &= |K_t^{c} \cap \phi_t^{c}| \\
    C_t &= \left (\sum_{n=1}^{n=N}\Gamma_c [n \in K_t] \right) ^ {1.5} \label{eq:C}\\
    H_t &= \sum_{n \in K_{t}} \sum_{v \in N^{+}(n)} (\vuln{v}[v \notin \phi_t])\\
    A_t &= \begin{cases}
        \Gamma_{\patch} & \mathcal{A}_t = \patch\\
        \Gamma_{\isolate} & \mathcal{A}_t = \isolate
    \end{cases}\\
    T_t &= C_t + A_t \label{eq:sem_finish}.
\end{align}

These structural equation models allow us to transfer observable information from \yt by manipulating properties of the environment so that we receive an expression for each variable in the \DAG. The time dependence of the two observable variables that have transition edges, namely $S_t$ and $C_t$, is implicit in the above equations. As $C_{t-1}$ increases (due to an increased number of compromised nodes), there is an implicit change in $K_t$ and therefore in $S_t$. Similarly, as $H_{t-1}$ increases due to more vulnerability in the network, there is an implicit increase in $C_t$ as additional nodes get compromised between times $t-1$ and $t$.

As noted, having access to observational data means we are able to estimate the \sem (since we do not have access to the true \sem):
\vspace{-0.25cm}
\begin{align*}
    P_t &= f_P(t) + \epsilon_P \\
    I_t &= f_I(t) + \epsilon_I \\
    S_t &= f_S(C_{t-1}, I_t) + \epsilon_S \\
    C_t &= f_C(H_{t-1}) + \epsilon_C \\
    H_t &= f_H(P_t, C_t) + \epsilon_H \\
    A_t &= f_A(P_t, I_t) + \epsilon_A \\
    T_t &= f_T(C_t, A_t) + \epsilon_T
\end{align*}

by placing Gaussian process estimators on all functions $f_i(\cdot), i \in \{P, I, S, C, H, A, T\}$. This \sem is then used to find the ground truth optimal intervention for the system at each time step, $\{y^\star_t \mid  t = 0,\dots, 24\}$, given it has full access to the data from \yt and the causal graph in \cref{fig:DAG}. 
\begin{figure*}[ht!]
    \centering
    \includegraphics[width=\textwidth]{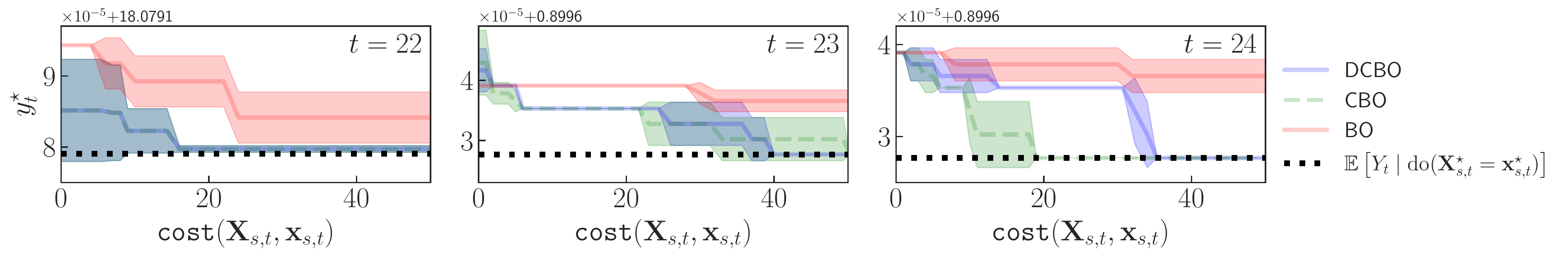}
    \caption{Experimental results of different optimisation methods applied to \yt data, showing convergence of \dcbo and competing methods (\cbo and \bo) across five replicates. The black dotted line shows the optimal response value $\{ y^*_t \mid  t=22,23,24 \}$. Shaded areas are $\pm$ one standard deviation. The $x$-axis shows the total \textit{cumulative cost} of the intervention set over 50 trials - effectively how many times the optimisation had to probe the underlying function in order to build up an approximation of the causal relationship.}
    \label{fig:dcbo_output}
\end{figure*}
\section{Experiments}
\label{sec:experiments}

To fit our initial \sem, we use \yt to generate observational data. This involves setting up a number of \yt environments each with a different dummy agent that has a probabilistic chance to take each action. We then step through the environments to see how well the chosen probabilities perform\footnote{Python code required to reproduce this experiment: \url{https://github.com/alan-turing-institute/causal-cyber-defence}}.

We configure a simple blue agent (\ba) that can take two actions, restore ($\mathcal{\patch}$) or isolate ($\mathcal{\isolate}$), with probabilities $p(\mathcal{\patch})$ and $p(\mathcal{\isolate})$. Additionally, if a node is isolated, it will automatically be reconnected to the network after five time steps.

We also configure a simple red agent (\ra) that can only perform a single action, a basic attack, that uses \cref{eq:basic_attack_1} and \cref{eq:basic_attack_2} with a skill level of $25\%$. A $25\%$ skill level is rather low and means that on average the weakest nodes will take four turns to compromise. The \ra chooses its target by picking a random node $V_i$ from the set of all nodes $V_i \in \Vnet$ such that $V_i$ is an entry node or $V_i$ is connected to a compromised node. In \cref{fig:YT_simulation} the \ra can target any node with a link to an orange node or any node marked with an `E'. In \cref{fig:YT_simulation}, the \ra has chosen a node in the bottom left. 
 
We initialise ten identical network environments with identical \ra's. We pick a simple network topology with ten network nodes. Combined with the basic action spaces of both the \ba and \ra, this simple setup allows the relatively simple \DAG in \cref{fig:DAG} to be constructed. Integrating \dcbo with more complex topologies and agents would be easy to do, but would require re-drawing the \DAG such that all causal relationships are accounted for. In the setup we randomly assign a node in the network to be the entry node into the network for the \ra. We then randomly select a single node to be a high value target (\hvt) from the set of nodes that are furthest away from the entry node -- the purple node in \cref{fig:YT_simulation}. The configuration selected for this experiment was chosen to create a non-trivial environment that may not necessarily reflect any real world systems but still contains features and challenges that agents would have to face in these systems. For each environment, we draw a single random value for $p(\mathcal{\patch})$ and $p(\mathcal{\isolate})$ from a Gaussian distribution centred around $0.5$, and instantiate a \ba using these values. We then run the simulation for 25 time steps, or until the \hvt is compromised, allowing both \ra and \ba to sequentially take actions according to their configurations. At each time step, we calculate the values for the nodes in \cref{fig:DAG} according to the \sem in \cref{eq:sem} to \cref{eq:sem_finish}.

This provides an array of observational data for each node of size $10 \times 25$. As in \citep[4.2 Real experiments]{dcbo}, \cbo and \dcbo use this data to fit a non-parametric simulation of the relationships in \cref{fig:DAG} and to compute a causal prior.

We provide both \cbo and \dcbo this observational data and run them, along with \bo, for 50 trials each, allowing them to intervene on the manipulative variables $P$ and $I$ in the domain $[0,1]$, to minimise $T$ at time steps $t=22,23,24$. The convergence of the models on the optimal outcome value is shown in \cref{fig:dcbo_output}, against the cost of intervention. Note that \cbo and \dcbo converge much more consistently and more efficiently than \bo.


\section{Results and discussion}
\label{sec:results_n_discussion}



\Cref{fig:dcbo_output} shows that \dcbo and \cbo provide a way of efficiently evaluating the optimal decisions to make in a complex cyber environment, with a time-dependent causal structure. They are able to take historic data, along with a causal model of the data generating process, and find an optimal intervention set, choosing actions with the least associated cost from both the action, and the effect of the \ra. 

A large part of both \cbo and \dcbo's efficiency, and an additional hurdle to implementing these causal methods in a real-world cyber setting, is the creation of the underlying \DAG on which these methods rely for inference. For a controlled setting, such as those within simulators such as \yt, the creation of this \DAG and corresponding observational data is relatively trivial. On the contrary, in a real-world setting, the \DAG creation must be knowledge-driven and can often grow to unmanageable scales when considering all the factors at play in a potentially hostile cyber environment. However, with sufficient domain expertise and an understanding of the data-generating process, we have shown that a causal view of activities within a complex cyber simulator is possible and indeed more powerful than a na\"ive data-centric view. We leave it as an exercise to the reader to think of \DAG's for real-world networks that they are familiar with, as we believe it can shed light on relationships that might otherwise go un-noticed.

The results show that \cbo generally converges to the optimal decision the quickest with \dcbo varying in its relative efficiency. This implies that in this specific network topology, the temporal relationships do little to aid in the convergence on the optimal solution. However, the causal relationships \textit{within} time slices, exploited by both \cbo and \dcbo, do significantly improve the convergence over traditional \bo. 

\begin{figure}[ht!]
\centerline{\includegraphics[width=\columnwidth]{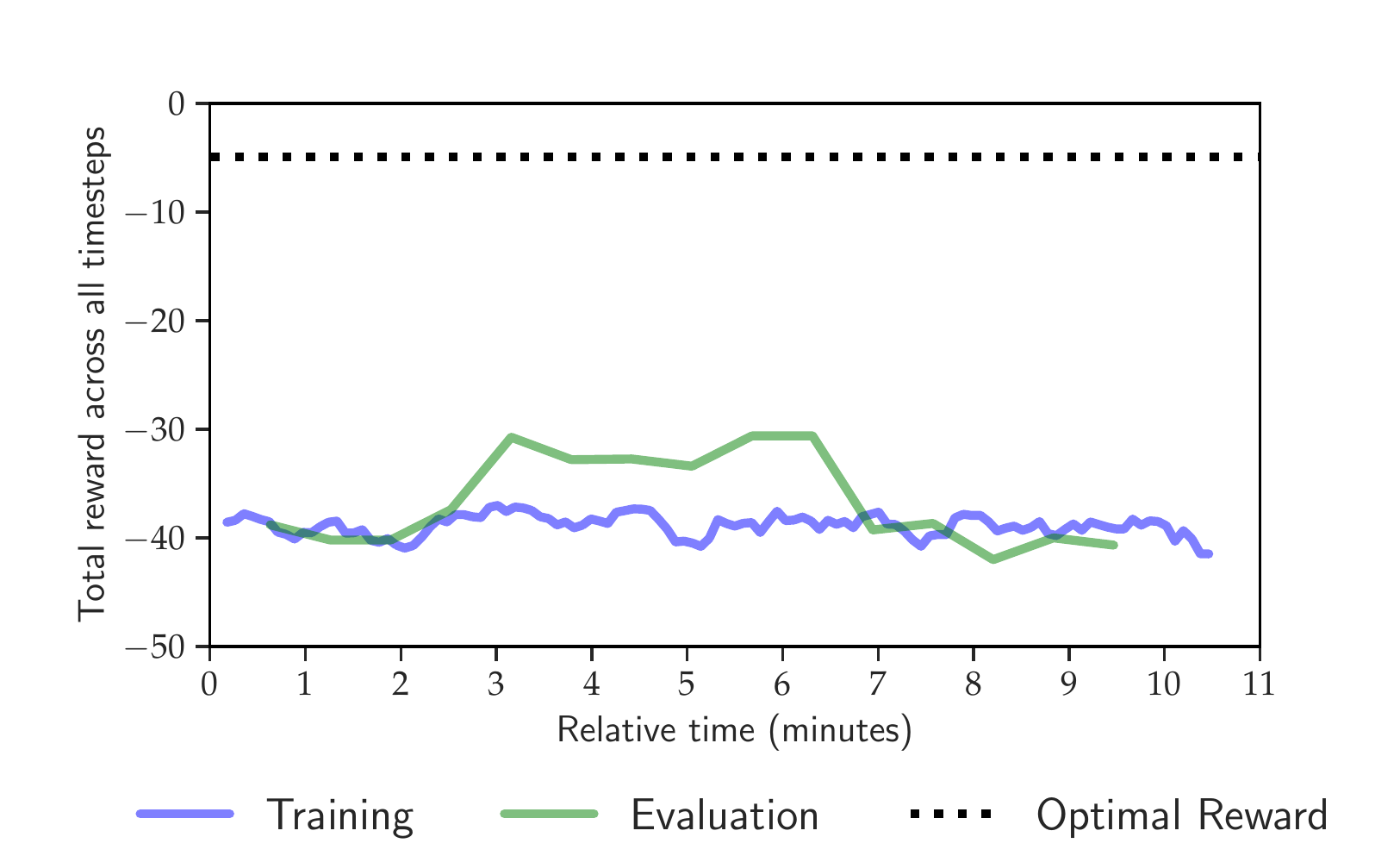}}
\caption{\ppo agent trained in the same environment as the other decision making algorithms. The reward on the horizontal axis is the cumulative reward over 25 time steps. The reward ($R$) at each timestep is $ R = - T $ where $T$ is the total cost calculated in \cref{eq:sem_finish}. The \ppo manages to reach its optimal solution within a couple of minutes but it should be noted that we encountered a considerable spread in performance when testing \ppo in this context.}
\label{fig:YT_training}
\end{figure}

One of the main advantages of \cbo and \dcbo is speed and their handling of data-sparsity. In addition, provided that the training environment remains broadly stationary, \dcbo and similar approaches are able to be rapidly re-trained and re-calibrate defensive policies as the threats and risk environment changes. Compare this to \rl where a major disadvantage of those methods is their sample inefficiency and training time. As a means of comparison, a Proximal Policy Optimisation (\ppo) agent \citep{schulman2017proximal} was trained using the stable-baselines3 \citep{stable-baselines3} \rl library within the same environment used to gather data for our comparison models. It took significantly longer to converge to its optimal interventions (see \cref{fig:YT_training}) and its optimal solution was far from the true optimal. This shows our possible need for another method to choose optimal decisions and \dcbo and \cbo have been shown to quickly hone in on the true solution. These results are not conclusive. We will need to perform further analysis with \rl to fully understand the pros and cons. 

\section{Conclusion and future work}
\label{sec:conclusion}


We have demonstrated the utility of using causal inference in a cyber-simulation framework where the goal is to provide a blue agent with optimal action recommendations. In that process we have made a number of operative assumptions, discussed throughout the paper. Due to the preliminary nature of the work, there are many avenues going forward.

On the causal inference side, we have assumed \emph{causal sufficiency} i.e. the absence of any unmeasured confounders. This is a strong and oftentimes unrealistic assumption. There is a host of literature which focuses on causal decision-making in the \rl domain, which we aim to bring across to the \bo sphere, such as the work by \citet{lee2018structural}. Moreover, it is equally true that we have only focused on continuous interventions in this paper but cyber-security data is often discrete, hence future work will incorporate options for handling discrete data.

On the \yt side, we hope to further explore causal decision making algorithms and their integration into \yt, increasing the complexity of the environments and number of actions available. We also hope to use these models in an on-line setting, where causal decision making is able to provide real-time recommendations of interventions, which are then fed directly into \yt's simulation.

With the on-line setting in place, we then plan to gather more results on how algorithms such as \dcbo fare against traditional reinforcement learning algorithms, such as \ppo, in these different and more complex environments. What is more, \dcbo is \emph{always} active but this may not be necessary. It would be better if \dcbo activated only \emph{if} a threat has been detected upon which an optimal causal decision-making agent is deployed. Early work on one such (non-causal) approach was recently published by \citet{hammar2022stopping} using optimal stopping \citep{bertsekas2012dynamic}. Finally, we hope that a systematic comparison of agents trained using a variety of different algorithms will show the value that a causal understanding of the cyber environment can bring to optimal cyber decision making.


\bibliographystyle{icml2022}
\bibliography{references}



\end{document}